\documentclass[prl,twocolumn]{revtex4}
\usepackage[dvips]{graphicx}
\usepackage{latexsym}
\begin{document}
\newcommand {\beq} {\begin{equation}}
\newcommand {\eeq} {\end{equation}}
\newcommand {\bqa} {\begin{eqnarray}}
\newcommand {\eqa} {\end{eqnarray}}
\newcommand {\rr} {\ensuremath{{\bf{r}}}}
\newcommand {\kk} {\ensuremath{{\bf{k}}}}
\newcommand {\no} {\nonumber}
\newcommand{\rh} {\ensuremath{\rho}}
\newcommand{\ep} {\ensuremath{\epsilon}}
\newcommand{\as} {\ensuremath{1/(k_Fa_s)}}
\newcommand{\kf} {\ensuremath{k_F^{-1}}}
\title{Vortices in Superfluid Fermi Gases Through the BEC to BCS Crossover}
\author{Rajdeep Sensarma, Mohit Randeria, and Tin-Lun Ho}
\affiliation{Department of Physics, The Ohio State University, Columbus, OH 
43210, USA}
\date{\today}
\begin{abstract}
We have analyzed a single vortex at $T=0$ in a 3D superfluid atomic Fermi gas  
across a Feshbach resonance. On the BCS side, the order parameter varies on two scales:
$k_{F}^{-1}$ and the coherence length $\xi$, while only variation on the scale of $\xi$ is seen 
away from the BCS limit. The circulating current has a peak value $j_{max}$ which is a non-monotonic function
of $1/k_F a_s$ implying a maximum critical velocity $\sim v_F$ at unitarity. The number of fermionic bound states in the core decreases 
as we move from the BCS to BEC regime. Remarkably, a bound state branch persists even on the BEC side
reflecting the composite nature of bosonic molecules.
 \vspace{0.1cm}
\typeout{polish abstract}
\end{abstract}

\maketitle

The recent discovery of vortices in $^{6}$Li  Fermi gases is a milestone in 
the development of fermion superfluid in atomic Fermi gases \cite{MITvortex}. For 
the first time, the phase coherence of 
a superfluid atomic Fermi gas is demonstrated unambiguously.  The 
atomic Fermi gases of alkali atoms, however, are very unusual many-body 
systems. Their interactions, which are related to s-wave 
scattering length, are highly tunable using Feshbach resonance.  By 
tuning the inverse scattering length $1/a_{s}$ continuously from  $-\infty$ 
to $+\infty$, the ground state of these systems changes from a weak coupling 
BCS superfluid to a molecular BEC \cite{leggett}.  The region of infinite scattering 
(or $a^{-1}_{s} =0$) is particularly interesting, where the Fermi gas 
exhibits universal behavior, in the sense that the interaction energy scale 
at $T=0$ is independent of atomic details and is given by Fermi energy
\cite{universal}, giving rise to a superfluid transition temperature $T_{c}$ comparable to Fermi temperature $T_{F}$.  
In fact, it is found that at resonance, $T_{c}/\ep_{F}\sim 0.2$\cite{Jin},  the highest of all known fermion superfluids.  

Although all vortices of s-wave BCS superfluids have topologically invariant properties like quantized circulation
of $h/2M$, other features like vortex core size, circulating current, and the bound state spectrum depend on dynamical details. It is natural to ask how the properties of a vortex change as one goes across the Feshbach resonance from the BCS to the BEC side, and how unitarity manifests itself in these properties. The purpose of this paper is to answer these questions using the Bogoliubov-deGennes (BdG)
approach \cite{Caroli} in a {\em three} dimensional system for a {\em wide} Feshbach resonance \cite{wide}. We shall show that at $T=0$: 

\noindent ${\bf (A)}$ On the BCS side, the order parameter $\Delta(\rho)$ for a single vortex exhibits \emph{two} length scales: an initial
rise on the scale of $k_F^{-1}$ \cite{comment1} for which we give an elementary analytical argument, and an eventual approach to its
bulk value $\Delta_{0}$ on the coherence length scale $\xi = \hbar v_{F}/\Delta_{0}$ where $v_{F}$ is the Fermi velocity.  At unitarity where $\Delta_{0} \sim \ep_{F} = \hbar^2 k_{F}^2/2M$, $\xi$ reduces to $k_{F}^{-1}$, and the two length scales coincide. On the BEC side 
the $\Delta(\rh)$ reaches its bulk value over the coherence length $\xi \sim 1/\sqrt{n a_{s}}$, where $n$ is the density.
We also find that the density $n(\rh)$ is depleted in the vortex core, and its value $n(0)$ at the center is dramatically reduced
as one goes from the BCS to the BEC limit.

\noindent ${\bf (B)}$ The circulating current $j(\rh)$ around the vortex core has a peak value $j_{max}$ which is non-monotonic across the resonance and reaches a maximum precisely at unitarity.  Its scale is set by 
the critical velocity $v_c$ which is determined by pair breaking on the BCS side, a {\em single-particle} effect, but by {\em collective} 
excitations on the BEC side. It is interesting to note that so far, $j_{max}$ is one of the very few properties of atomic Fermi gases we know 
that varies non-monotonically across resonance, in contrast to all thermodynamic properties which vary monotonically and smoothly.

\noindent ${\bf (C)}$ We find that unitarity represents the most robust superfluid state in the entire BCS-BEC crossover.
Not only does one obtain the highest $T_c$ at unitarity (which is, however, not too different
from the $T_c$ value \cite{Randeria1} for all $1/k_F a_s >0$), but one also obtains the highest critical velocity $v_c \sim v_F$.

\noindent ${\bf (D)}$ We find that as we go from the BCS to the BEC limit, the number of fermionic bound states in the vortex core decreases, 
with a corresponding increase in both the energy of the lowest bound state and their level spacing. Remarkably, we find that a bound
state is observed way past unitarity, deep into the bosonic regime, which is unique to the molecular BEC. We also find that motion along
the vortex core broadens a bound state of angular momentum $\ell$ into a band with $k_z$ dispersion.

The BdG approach was first used to study vortices in BCS superconductors in the classic work of Caroli, de-Gennes 
and Matricon \cite{Caroli}; see also \cite{Schlutter}. In the superfluid atomic Fermi gases, the BdG 
approach in two dimensions has been used in \cite{Bruun} in the BCS limit and in \cite{Machida}. The latter work is
implicitly restricted to ``narrow resonances", which has very different physics
from the wide resonances relevant to all current experiments \cite{wide,Bulgac-comment}. 
The vortex problem has also been studied using density functional \cite{Bulgac}
and hydrodynamic \cite{Castin} approaches, which are not microscopic and hence do not provide information 
such as ${\bf (A)}$ to ${\bf (D)}$ mentioned above. 

{\bf Bogoluibov-deGennes (BdG) approach:}  For a wide resonance, it is sufficient to use the single channel model \cite{wide}. The BdG equations in this case are
\begin{equation}  
\left( \begin{array}{cc} \hat{T} & \Delta(\rr) \\ \Delta^{\ast}(\rr) & - \hat{T}^{\ast}\end{array}\right)
\left( \begin{array}{c} u_n (\rr) \\ v_n (\rr) \end
{array} \right) = E_n
\left( \begin{array}{c} u_n (\rr) \\ v_n (\rr) \end
{array} \right) 
\label{BdG} \end{equation}
where $\hat{T}= -\hbar^2\nabla^2/2M - \mu$, $\mu$ is the chemical 
potential, $E_n$ are the eigenvalues,  $u_n$ and $v_n$ are the 
eigenfunctions normalized by
\beq
\int d^3\rr  \left[ u_m^{\ast}(\rr)u_n(\rr)+v_m^{\ast}(\rr)v_n(\rr)\right]=\delta_{mn}
\label{norm}
\eeq
The order parameter $\Delta(\rr)$ and the chemical potential $\mu$ are determined by the self-consistency equation 
$ \Delta(\rr) = g\sum_n u_n(\rr) v_n^{\ast}(\rr)$ and the average density $n=2\sum_n\int d^3\rr |v_n(\rr)|^2$. The $\sum_n$ is restricted to $0\leq E_n\leq E_c$ where $E_c$ is a high energy cutoff (discussed below).
>From now on we measure all energies in units of  $\ep_F$ and lengths in units of $k_F^{-1}$, so that the bare attraction $g$ has units of $k_F^3/\ep_F$. The cutoff $E_c$ and the corresponding $g(E_c)$ should be chosen to satisfy 
\begin{equation}
\frac{1}{k_Fa_s }= -\ \frac{8\pi\ep_F}{gk_F^3}+\frac{2}{\pi}\sqrt{\frac{E_c}{\ep_F}}
\label{asc}  \end{equation}
so that the low energy effective interaction is described by the scattering length $a_s$ \cite{Randeria1}.

We work in cylindrical co-ordinates $\rr=(\rh,\theta,z)$ in a gauge in which $\Delta(\rr )=\Delta(\rh)e^{-i\theta}$. Working in a cylindrical box of radius $R$ and height $L$, our normalized wavefunctions are of the form 
$u_n(\rr)=u_n(\rh)e^{i\ell\theta}e^{ik_zz}/{\sqrt{2\pi L}}$ and
$v_n(\rr)=v_n(\rh)e^{i(\ell+1)\theta}e^{ik_zz}/{\sqrt{2\pi L}}$ 
so that (\ref{BdG}) decouples into different $l$ and $k_z$ sectors \cite{Schlutter}. We further expand the radial wavefunctions $u_n(\rh)=\sum_j c_{nj}\phi_{j\ell}(\rh)$ and $v_n(\rh)=\sum_j d_{nj}\phi_{j\ell+1}(\rh)$ in the orthonormal basis set
$\phi_{j\ell}(\rh)=\sqrt{2}J_{\ell}(\alpha_{j\ell}\rh/R)/[RJ_{\ell+1}(\alpha_{j\ell})]$
 where $\alpha_{j\ell}$ is the $j^{th}$ zero of $J_{\ell}(x)$. The BdG eigenvalue equation now reduces to a matrix diagonalization problem
\begin{equation}  
\left( \begin{array}{cc} T_{\ell} & \Delta_{\ell} \\ \Delta_{\ell+1} & - T_{\ell+1}\end{array}\right)_{jj'}
\left( \begin{array}{c} c_{nj'} \\ d_{nj'} \end
{array} \right) = E_n
\left( \begin{array}{c} c_{nj}  \\ d_{nj}  \end
{array} \right) 
\label{matrix}
\end{equation}
\begin{figure}[t!]
\includegraphics[scale=0.3,angle=270]{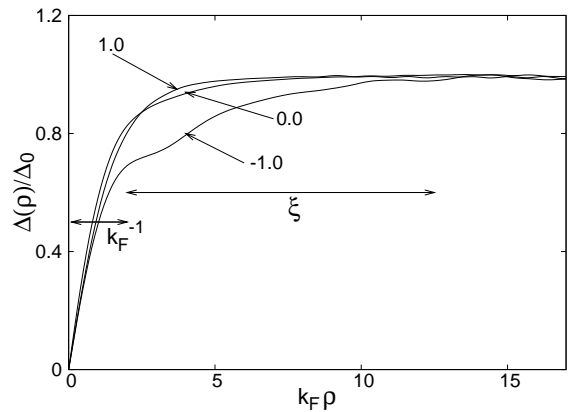}
\caption{The order parameter normalized by its value far from the vortex for three different couplings
$1/k_F a_s = -1, 0, 1$. Note the clear emergence of two length scales in the BCS limit $1/k_F a_s = -1$.}
\label{lengthscale}
\end{figure}

\noindent where $T_{\ell}^{jj'}=(\alpha_{j\ell}^2/R^2+k_z^2-\mu)\delta_{jj'}$ and $\Delta_{\ell}^{jj'}=\int d\rh \Delta(\rh)\phi_{j\ell}(\rh)\phi_{j'\ell+1}(\rh)$  
While the different $\ell$ and $k_z$ sectors are completely decoupled in (\ref{matrix}), they are coupled through the self-consistency equations \cite{details}.  Since the BdG equations are invariant under  $E_n\rightarrow -E_n$, $u_n(\rr)\rightarrow v_n^{\ast}(\rr)$, $v_n(\rr)\rightarrow -u_n^{\ast}(\rr)$, one can get the positive energy eigenfunctions for negative $\ell$ by looking at the negative energy eigenfunctions for positive $\ell$. This reduces the computational effort by half.

Ideally one would like to take $E_c \rightarrow \infty$ and obtain solutions independent of this cutoff.
In practice the size of Hilbert space grows like $RL\sqrt{E_c}$ and to make the calculation
manageable we choose $E_c=9\ep_F$, $R=25k_F^{-1}$ and $L=10k_F^{-1}$. We have checked that for this choice
of cutoff, which is already larger than that in \cite{Machida}, our results in the uniform case are no more than $5\%$ different from the infinite cutoff answers at unitarity. 

{\bf Vortex core structure:}
The order parameter $\Delta(\rh)$ is plotted in Fig.~1 for various values of $1/k_F a_s$. 
The weak oscillations in $\Delta(\rh)$, most prominent in the BCS regime, are likely finite size effects with no physical significance \cite{oscillations}. Another very interesting aspect of the BCS regime, clearly visible in  in fig.~\ref{lengthscale},
is the presence of {\it two} length scales $k_F^{-1}$ and $\xi$ in
the $T=0$ result for $\Delta(\rh)$. This is in marked contrast to the Ginzburg-Landau result
where $\Delta_{\rm GL}(\rh) \sim \tanh(\rho/\xi)$. While this effect was, in fact, recognized in the early
superconductivity literature \cite{KP} in an Eilenberger calculation, we provide an elementary analytical derivation here.
Close to the origin, one may ignore $\Delta$ relative to the kinetic energy terms in the BdG equations
and find $u_{\ell} = A_{\ell}^{-1} J_{\ell}(k_{F}\rho)$, $v_{\ell}= A_{\ell}^{-1} J_{\ell+1}(k_{F}\rho)$
where $A_{\ell}$ are constants.  For small $\rho$ we thus find $\Delta(\rho) = g u_{0}(\rho)v_{1}(\rho) = A_{0}^{-2}k_{F}\rho$.  
We next determine $A_{0}$ as follows. For large $\rh$,  $\Delta(\rh)=\Delta_0$ and $u_0 \sim (1/\sqrt{B_0\rh})\cos(k_F\rh)exp(-\rh/\xi)$ 
and $v_0=(1/\sqrt{B_0\rh})\sin(k_F\rh)exp(-\rh/\xi)$ (upto irrelevant phase shifts) where $B_0$ is another constant. 
Matching the large and small $\rh$ solutions at $\rh=\xi$, we get $A_0 \sim (B_0\kf)^{1/2}$ \cite{normalization}. 
Finally using the normalization condition (\ref{norm}) we can fix the constant $A_0=\xi/(k_F\xi)^{\frac{1}{2}}$ 
which leads to the result $\Delta(\rh) \sim \Delta_0 k_F \rh$ for small $\rh$. 
Thus in the BCS limit, the initial slope of $\Delta(\rh)$ is set by $k_F^{-1}$ although the eventual approach to 
its uniform value $\Delta_0$ is on a second scale of $\xi=\hbar v_F/\Delta_0 \gg k_F^{-1}$. 

\begin{figure}[t!]
\includegraphics[scale=0.35,angle=270]{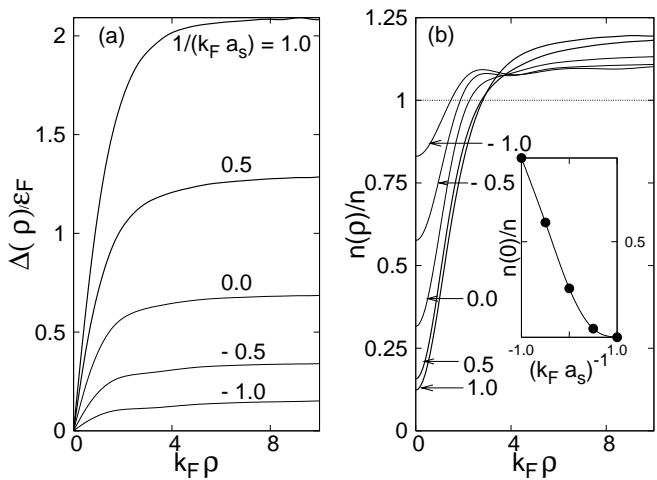}
\caption{(a) The order parameter profile and (b) the density profile for different values of $\as$.  Inset: the density at the center of vortex
as function of $\as$. }
\label{profile}
\end{figure}

We see that outside of the BCS regime there appears to be a single length scale in $\Delta(\rh)$ as seen in Fig.~1.
As the coupling increases toward unitarity the order parameter value at large $\rh$, $\Delta_{0}$ increases toward $\ep_F$
and the scale $\xi$ shrinks to $\sim k_F^{-1}$. 
The overall behavior of $\Delta(\rh)$ across the resonance is shown in Fig.~\ref{profile}(a). At the same time the density profile
$n(\rh)$ around the vortex evolves as shown in Fig.~\ref{profile}b. Near the center of the vortex 
$n(\rh) \simeq n(0)+a\rh^2$. The density at the center $n(0)$  is a strongly decreasing function of
$1/k_F a_s$ as shown in the inset of fig.~\ref{profile}(b) dropping from approximately 
$0.8 n$ at $1/k_F a_s = -1$, corresponding to a nearly ``full'' core, to $0.1 n$ at $1/k_F a_s = +1$, which is
a nearly ``empty'' core. 
  
{\bf Circulating current:}  The current circulating around the vortex core is ${\bf j}= \rh_s {\bf v}_s$, where 
$\rh_{s} \sim \Delta^2$, and ${\bf v}_{s} = (\hbar/2M\rh) \hat{\theta} $.  Far away from the vortex core,
$j \sim 1/\rh$ since $\Delta \rightarrow \Delta_{0}$.  On the other hand, near the center of the vortex, 
$\Delta \sim \rho$, so that $j \sim \rh$.  This implies that the current must have a peak $(j_{max})$ at a distance $\rh^{\ast}$ 
from the center,  and $\rh^{\ast}$ can be taken as a measure of the size of the vortex core. 
Intuitively, as the current increases with decreasing $\rh$ it eventually
reaches the \emph{critical current} at $\rh^{\ast}$, so that for $\rh < \rh^{\ast}$ the superfluid order parameter
is destroyed. 

Formally the current density is given by
\beq
{\bf j}(\rh)=-\frac{\hbar}{m\rh}\sum_{nlk_z}(l+1)[ \sum_{j}d_{nj}\phi_{jl}(\rh) ]^2 \hat{\theta}.
\eeq 
\begin{figure}[h!]
\includegraphics[scale=0.35,angle=270]{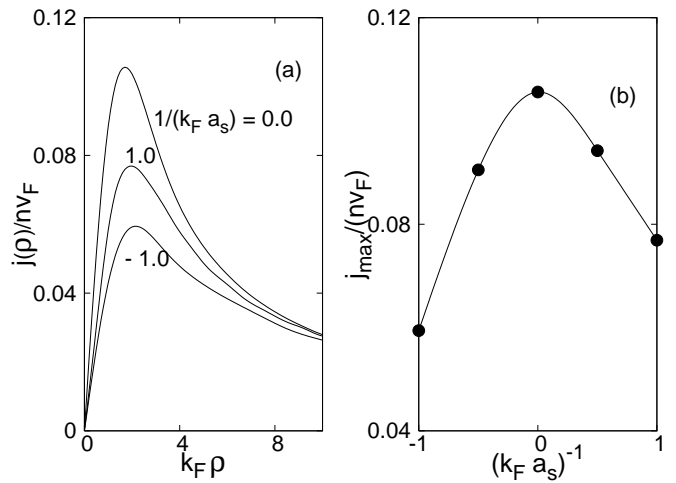}
\caption {(a) The current distribution for three different values of $\as$ (b) The peak current $j_{max}$ as function of $\as$. }
\label{current}
\end{figure}
The profiles of $j(\rho)$ for various $1/k_{F}a_{s}$ are shown in Fig.~\ref{current}(a). They all show a peak, as explained above.
We find that the location $\rh^{\ast}$ of the maximum current shows a weak non-monotonic behavior 
in the vicinity of unitarity, which is roughly consistent with the non-monotonic behavior of the coherence
length $\xi$ as a function of $1/k_F a_s$ predicted earlier; see Fig.~3 of \cite{Randeria2}. Details will be published separately
\cite{sensarma}. 

The peak current $j_{max}$, however, shows a non-monotonic behavior as a function of $1/k_{F} a_{s}$, peaking at unitarity
as shown in Fig.~\ref{current}(b). We may understand this non-monotonic behavior as follows.
On the BCS side, the critical velocity, determined by depairing, is $\Delta_0/k_F$ and thus $j_{max} \sim n v_{F} (\Delta_{0}/\epsilon_{F})$.
The critical current thus increases as one moves toward unitarity from the BCS side as $\Delta_0$ increases.  
At unitarity, universality dictates that the critical velocity must be of order $v_F$ so that $j_{max} \sim n v_F$.
As one goes toward the BEC side the critical current is now determined not by pair breaking but rather by the collective
excitations. The Landau criterion suggests $v_c \sim v_F\sqrt{k_F a_s}$ which leads to a $j_{max}$ which decreases
with increasing $1/k_F a_s$. We thus find the interesting result that
the mechanism for destruction of superfluidity, as reflected in the maximum current $j_{max}$, changes as one goes across the resonance
from pair breaking on the BCS side to collective on the BEC side. 

{\bf Spectrum:}
\begin{figure}[t!]
\includegraphics[scale=0.35,angle=270]{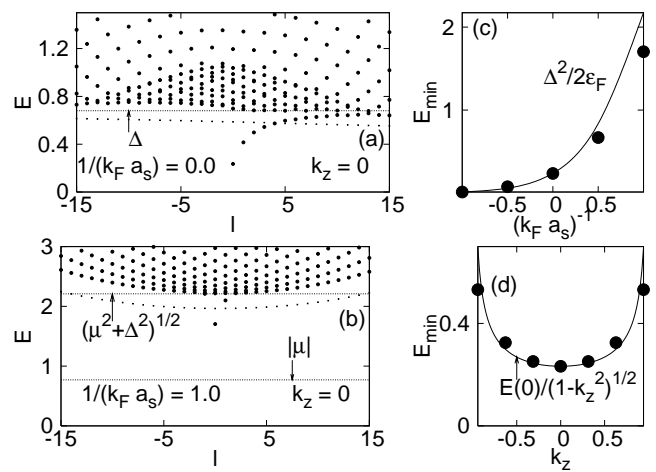}
\caption{The BdG spectrum in the $k_z=0$ sector as a function of $\ell$ for (a) $\as=0.0$, i.e., unitarity and (b) $\as=1.0$, i.e., on BEC side. (c) The minigap ($\ell=0$ $k_z=0$) as a function of $\as$. (d) The $k_z$ dependence of the bound state energy for $\ell=0$ at unitarity.}
\label{spectrum}
\end{figure} 
\noindent
We now turn to the fermionic bound states in the vortex core and their evolution through the BCS-BEC crossover.
In the BCS limit the results are very well known from CDM \cite{Caroli}.
They showed that for each $\ell$ and $k_z$ there is a bound state,
with energy less than $\Delta_0$, in the core of the vortex. The lowest energy fermionic excitation 
(with quantum numbers $\ell=0$ and $k_z=0$) has a ``minigap'' $\Delta_0^2/2\ep_F \ll \Delta_0$.
As we increase $1/k_F a_s$ we see some changes in the spectrum. Our results at unitarity $1/k_F a_s = 0$
and in the BEC regime $1/k_F a_s = 1$ are shown in Fig.~\ref{spectrum} (a) and (b) respectively.
For clarity we show the spectrum only as a function of angular momentum $\ell$ and at fixed $k_{z}=0$
($k_z$-dispersion is discussed later).

As long as the chemical potential $\mu > 0$ (which includes unitarity), the bound (continuum) 
states are those with energies smaller (larger) than $\Delta_{0}$.
The bound state spectrum at unitarity (Fig.~\ref{spectrum}(a)) is not qualitatively different from the BCS limit, except that both
the minigap and the level spacing are larger, and therefore one has fewer bound states. Remarkably, the minigap continues to follow
$\Delta_0^2/2\ep_F$ even through unitarity as shown in Fig.~\ref{spectrum}(c).

Once the chemical potential $\mu < 0$, as in the BEC regime shown in Fig.~\ref{spectrum}(b), the continuum of fermionic excitations 
exists for $E \geq (|\mu|^2+\Delta_0^2)^{\frac{1}{2}}$ \cite{leggett}. We can still define a fermionic bound state by demanding that 
the corresponding wavefunction decays exponentially to zero away from the vortex core. If such states exist, then it can be easily 
shown that their energy must lie in the interval $|\mu| \leq E < (|\mu|^2+\Delta_0^2)^{\frac{1}{2}}$.
Remarkably, we find such a bound state well into the BEC regime as evident from Fig.~\ref{spectrum}(b).
It is amusing that the \emph{off-diagonal} potential $\Delta(\rh)$ produces an Andreev \emph{bound} state even at an energy 
larger than the maximum value of the potential $\Delta_0$, but we must emphasize that this bound state \emph{is} below the continuum 
which starts at $(|\mu|^2+\Delta_0^2)^{\frac{1}{2}}$.
The fermionic bound states are a unique consequence of composite nature of bosons in the molecular BEC and are absent in atomic BEC. 

As a result of motion along the vortex axis, each state for fixed angular momentum $\ell$ shown in Fig.~\ref{spectrum} (a) and (b) 
actually broadens into an energy band with $k_z$-dispersion. Our calculated $k_z$ dependence of the bound state energy for the $l=0$ bound state at unitarity is shown in Fig.~\ref{spectrum}(d). The energies (which are discrete due to the finite $L$ in the z-direction)
continue to follow the Caroli et. al. \cite{Caroli} BCS-limit prediction  $E_0/(1-k_z^2)^{\frac{1}{2}}$ even at unitarity .
 
Finally, we note that there is a branch of bound states (shown with smaller dots in Figs.~\ref{spectrum}(a) ~\ref{spectrum}(b))
which lies below the fermionic energy gap in the bulk but differs from the vortex core states (which exist only for $\ell \geq 0$)
in that it exists both for positive and negative $\ell$. From their wavefunctions we deduce that these states are not related to the vortex core, 
but are in fact trapped near $\rho=R$ due to the suppression of the order parameter by the hard wall boundary  condition \cite{ohashi}. 
The details of these as well as the vortex bound state wavefunctions will be described separately \cite{sensarma}.

{\bf Conclusions :}  The physics of a vortices in the BEC to BCS crossover has led to interesting results 
$({\bf A})$ through $({\bf D})$ summarized in the introduction.  The study of the order parameter profile in the vortex core and the circulating current may require new experimental methods which will allow more detailed diagnostics. The existence of fermionic bound states should be detectable by spectroscopic means, or through the damping of sound, or through effects of dissipation in vortex dynamics.  
We hope that the unique properties of vortices that we point out here will stimulate further experimental and theoretical studies on new methods to probe strongly interacting Fermi gases across Feshbach resonance. 

This work is supported by NASA GRANT-NAG8-1765 and NSF Grant DMR-0426149.


\begin{thebibliography}{99}
\bibitem{MITvortex} 
M.W. Zwierlein, et.al, Nature {\bf 435}, 1047 (2005).
\bibitem{leggett}
A. J. Leggett in {\it Modern Trends in the Theory of Condensed Matter},
A. Pekalski and R. Przystawa, eds. (Springer-Verlag, Berlin, 1980). 
\bibitem{universal} 
See K. O'Hara {\it et al.}, Science {\bf 298}, 2179 (2002); 
T.L. Ho, Phys. Rev. Lett. {\bf 92} 090402 (2004), and references therein.
\bibitem{Jin}  
C. A. Regal, {\it et al.},  Phys. Rev. Lett. {\bf 92}, 040403 (2004); 
M. W. Zwierlein, {\it et al.}, Phys. Rev. Lett. {\bf 92}, 120403 (2004). 
\bibitem{Caroli} 
C. Caroli, P. de Gennes and J. Matricon, Phys. Lett. {\bf 9},307 (1964).
\bibitem{wide} 
That current experiments are for wide resonances is emphasized in R. Diener and T.L. Ho, cond-mat/0405174. See also D.S. Petrov 
{\it et al.}, cond-mat/0502010. 
\bibitem{comment1} 
While it was recognized in earlier work \cite{KP,Schlutter}, this fact nevertheless does not seem to be widely appreciated.
\bibitem{KP} 
L. Kramer and W. Pesch, Z. Phys. {\bf 269}, 59 (1974). 
\bibitem{Schlutter} 
F. Gygi and M. Schlutter, Phys. Rev. B. {\bf 43}, 7609 (1991).
\bibitem{Randeria1} 
C. A. R. Sa de Melo, M. Randeria and J. R. Engelbrecht, Phys. Rev. Lett. {\bf 71}, 3202 (1993).
\bibitem{Bruun}  
N. Nygaard, {\it et al.}, Phys. Rev. Lett. {\bf 90}, 210402 (2003).
\bibitem{Machida} 
M. Machida and T. Koyama,  Phys. Rev. Lett. {\bf 94}, 140401 (2005).
\bibitem{Bulgac-comment} 
A. Bulgac, cond-mat/0505524. 
\bibitem{Bulgac} 
A. Bulgac and Y. Yu,  Phys. Rev. Lett. {\bf 91}, 190404 (2003.
\bibitem{Castin} 
G. Tonini and Y. Castin, cond-mat/0504612. 
\bibitem{details} 
The self consistency equations now reduce to $\Delta(\rh)=(g/2\pi L)\sum_{njj'lk_z}c_{nj}d_{nj'}\phi_{jl}(\rh)\phi_{j'l+1}(\rh)$ and 
$n=(2/\pi R^2L)\sum_{njlk_z}d_{nj}^2$.
\bibitem{oscillations} 
We believe that the completely smooth $\Delta$ profile seen in \cite{Schlutter}, which used a simple BCS interaction and a very large system size, shows that the oscillations seen in the BCS limit of our dilute gas model are a finite size artifact, and not Friedel like oscillations 
as suggested in \cite{Bruun,Machida}.
\bibitem{normalization} Equating $A_{0}^{-1}J_{0}(k_{F}\xi) = (B_{0}\xi)^{-1/2}{\rm cos}(k_{F}\xi) e^{-1}$, and using $J_{0}(x)\sim \cos x /\sqrt{x}$ for $x \gg 1$, we have $A_{0} \sim (B_{0}k_{F}^{-1})^{1/2}$.
\bibitem{Randeria2}
J.R. Engelbrecht, M. Randeria and C.A.R. Sa de Melo, Phys. Rev. B {\bf 55}, 15153 (1997).
\bibitem{sensarma}
R. Sensarma, M. Randeria and T. L. Ho, (unpublished).
\bibitem{ohashi}
Y. Ohashi and A. Griffin, Phys. Rev. A {\bf 72}, 013601 (2005)


\end{thebibliography}
\end{document}